\documentclass[amsmath,amssymb,aps,prl,reprint,floatfix,superscriptaddress]{revtex4-2}

\usepackage{graphicx}
\usepackage{dcolumn}
\usepackage{xcolor}
\usepackage{bm}
\usepackage[version=4]{mhchem}
\usepackage[breaklinks,colorlinks,linkcolor={blue}, %
            citecolor={blue},urlcolor={blue}]{hyperref}
\usepackage[normalem]{ulem} 
\begin{document}

\title{Do Small Hole Polarons form in Bulk Rutile TiO\texorpdfstring{$_2$}{2}?}
\author{Shay McBride} 
\affiliation{Thayer School of Engineering, Dartmouth College, Hanover, New Hampshire 03755, USA}
\author{Wei Chen}
\affiliation{Institute of Condensed Matter and Nanoscicence (IMCN),
Universit\'{e} catholique de Louvain,
Louvain-la-Neuve 1348, Belgium}
\author{Tanja \'Cuk}
\affiliation{University of Colorado, Boulder, Colorado 80309, USA}
\author{Geoffroy Hautier} 
\affiliation{Thayer School of Engineering, Dartmouth College, Hanover, New Hampshire 03755, USA}
\date{\today}

\begin{abstract}

Hole transport and localization through small polarons is essential to the performance of TiO$_2$ in photocatalysis applications. The existence of small hole polaron in bulk rutile TiO$_2$ has been however controversial with contradicting evidences from theory and experiments. Here, we use first principles computations and more specifically a Koopmans' compliant hybrid functional and charge correction to study small hole polarons in bulk rutile. We find that a fraction of exchange exists satisfying Koopmans' compliance for the polaron state and reproducing the band gap provided that charge correction is used. We clearly show that first principles computations indicate that the small hole polaron is unstable in bulk for rutile and stable for anatase TiO$_2$. 

\end{abstract}

\maketitle



Titanium dioxide (TiO$_2$) is a widely studied wide band gap semiconductor, known for its diverse range of applications, particularly in catalysis. TiO$_2$ is renowned for being the most generally used photocatalyst, owing to its abundance and high chemical stability \cite{scire_catalytic_2021}. A crucial factor in determining and understanding its effectiveness in catalytic and photocatalytic applications is the behavior of charge carriers, especially their transport mechanisms. 

The study of carrier transport in TiO$_2$ has garnered significant attention, particularly in the context of polaron formation and dynamics. In polarizable materials such as TiO$_2$, polarons, quasiparticles consisting of an excess charge coupled with lattice vibrations, play a pivotal role in dictating the optical and electronic properties. Of specific interest are hole polarons, which are essential to the oxidation process during photo-electrocatalysis and have become a focal point of research in recent years with important advances in their characterization \cite{cheng_hole_2012,lyle_electron-transfer_2021}. 

While the electron polaron in rutile TiO$_{\rm 2}$ is well-characterized both theoretically \cite{deskins_electron_2007, de_lile_polaron_2022,janotti_dual_2013} and experimentally \cite{tanner_photoexcitation_2021,franchini_polarons_2021}, the nature of the hole polaron in bulk rutile TiO$_{\rm 2}$ remains contentious. Conflicting theoretical reports in the literature, coupled with a dearth of experimental evidence, have fueled this ambiguity. 

Both experiment and theory agree that excess electrons in rutile TiO$_{\rm 2}$ are bound as reduced Ti$^{3+}$, but there is no consensus on trapped holes. Possible trap states for small hole polarons include bulk lattice O$^{\bullet-}$ anions, surface O$^{\bullet-}$ anions, and OH$^\bullet$ radicals. 
Experimentally, electron paramagnetic resonance, transient absorption spectroscopy, and photoluminescence spectroscopy have been used to study the trapping of holes in TiO$_{\rm 2}$, but the difficulty in distinguishing between polaron configurations makes the experimental picture unclear. Many spectroscopic studies probe holes to explore surface phenomena and transport on TiO$_{\rm 2}$ \cite{nakamura_molecular_2005,imanishi_atomic-scale_2014,kafizas_where_2016, amano_trapping_2024}. These studies cannot distinguish between surface and bulk trap states, and offer no definitive conclusion about the bulk hole polaron. Electron paramagnetic resonance studies reached opposite conclusions on the presence of hole polarons in rutile TiO$_{\rm 2}$\cite{yang_photoinduced_2010, macdonald_situ_2010}. Photoluminescence studies have assigned the 840 nm emission peak in rutile to both the deep electron trap \cite{wang_trap_2010,rex_spectroelectrochemical_2016} or a deep hole trap at O$_{\rm 3c}$ \cite{nakamura_molecular_2005,nakamura_primary_2004}.

Theoretical works are equally split, with studies as recent as this year reporting stable small hole polarons in bulk TiO$_{\rm 2}$\cite{amano_trapping_2024}. Even studies using the same hybrid functional parameters (HSE06) have reached opposite conclusions on the existence of the bulk hole polaron \cite{deak_polaronic_2011, cheng_identifying_2014, kokott_first-principles_2018}. Additionally, many theoretical studies on hole polarons in rutile TiO$_{\rm 2}$ make no mention of the bulk polaron at all, reporting formation energies only for polarons that form at surface oxygens, perhaps due to the confusion in the literature over the existence of bulk polarons.

The modeling of polarons has made much progress over the last decade. Semilocal functionals in density functional theory (DFT) generally fails to describe polarons because of their spurious electron self-interaction. Hybrid functionals are a way to address self-interaction, which manifests through the deviation from piecewise linearity of the total energy upon electron occupation. Recent works have shown that adjusting the hybrid functional parameters to enforce this piecewise linearity yields through the Koopmans' conditions can provide highly accurate electronic structures that agree with GW computations \cite{elmaslmane_first-principles_2018,miceli_nonempirical_2018,deak_choosing_2017,sadigh_variational_2015,chen_nonunique_2022,wing_role_2020}. Additionally, many correction schemes have been devised to correct for short- and long- range charge and polarization effects in defect and polaron supercell computations \cite{kokott_first-principles_2018, falletta_finite-size_2020}. 

Here, we show that careful computations with hybrid functionals tuned to enforce the Koopmans' condition and appropriate charge corrections clearly show that the hole polaron is not stable in bulk TiO$_{\rm 2}$ rutile. Our work resolves the current controversy and highlights the importance of Koopmans' condition and charge correction approaches in modeling polarons.

\begin{figure*}[t!]
 	\centering
 	\includegraphics[width=0.8\textwidth]{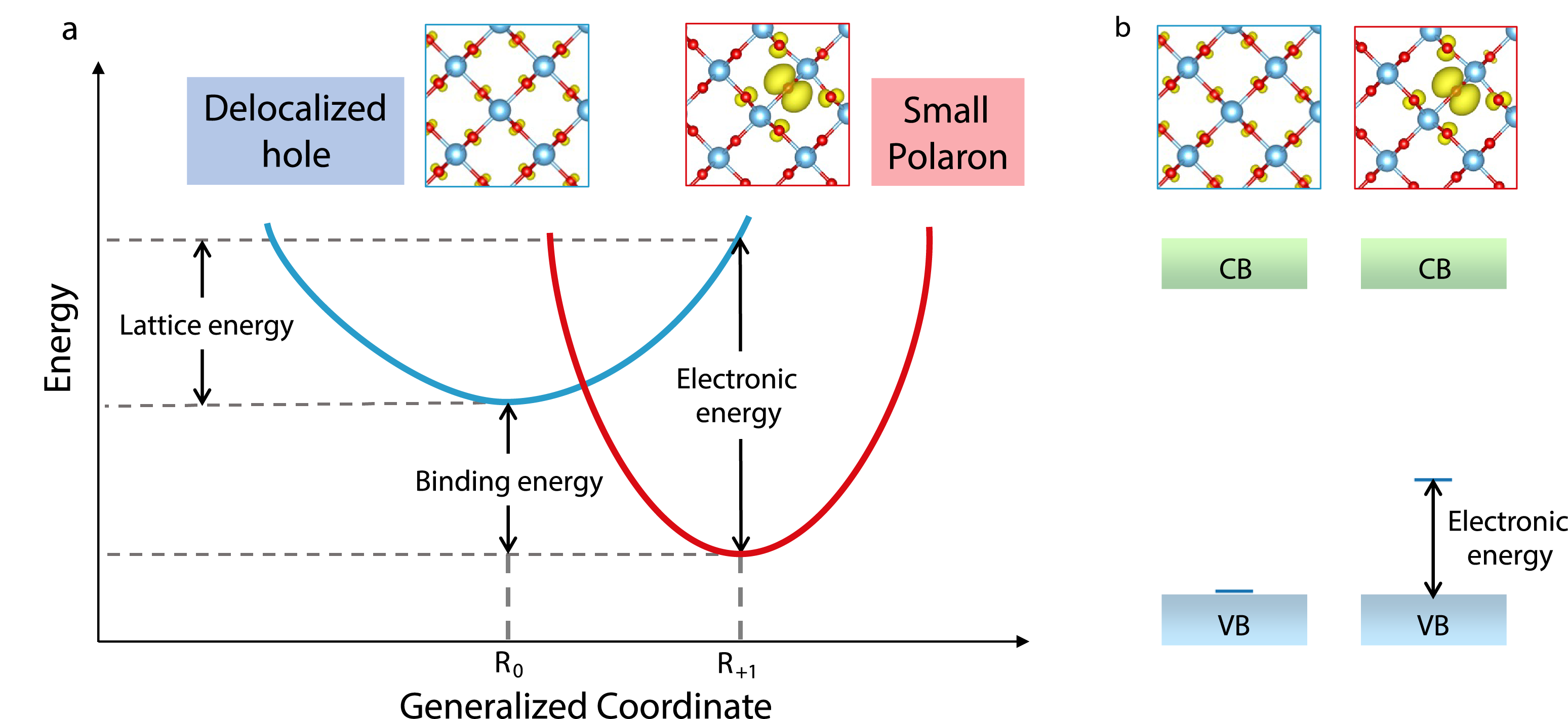}    
 	\caption{(a) Configuration coordinate diagram of energy as a function of lattice distortion for a delocalized hole and a localized small polaron. (b) Charge density isosurfaces of delocalized hole (left) and small polaron (right) in rutile TiO$_{\rm 2}$ and the corresponding mid-gap state.}
 	\label{Fig.1 polaron diagram}
\end{figure*}\textbf{}

A small polaron is formed when the interaction between an excess charge and lattice is strong enough to localize the charge within a small region, distorting the lattice and creating a potential well (Fig.~\ref{Fig.1 polaron diagram}a). We can model self-trapped polarons in a charged supercell by initializing a small distortion around an atom, then allowing the atoms to relax. The localized charge creates a state in the gap that affects the optoelectronic properties of the material (Fig.~\ref{Fig.1 polaron diagram}b). 

Density functional theory (DFT) with conventional local or semilocal functionals is not sufficient to describe polarons, and is well known to underestimate the band gap of transition metal oxides. Many approaches have been proposed to overcome this issue, such as DFT+U or hybrid functionals which include a fraction of Fock exchange ($\alpha$) \cite{franchini_polarons_2021, de_lile_polaron_2022}. However, these calculations depend on the choice of Hubbard $U$ correction or $\alpha$. Polarons are extremely dependent on correctly mixing Fock exchange and DFT exchange, and are in fact often used to parameterize hybrid functionals \cite{chen_nonunique_2022, wing_role_2020}.

We use the PBE0 hybrid functional which mixes the semilocal GGA-PBE functional with a fraction of Fock exchange $\alpha$ \cite{guidon_robust_2009}. We tune the $\alpha$ from first principles by enforcing the generalized Koopmans' condition. Koopmans' condition is an exact physical constraint based on the fact that total energy shows a piecewise linear dependence on addition of electrons. For a localize system, the ionization energy of system with $N$ electrons is identical to electron affinity of system with $N-1$ electrons, i.e.\ a perfect functional will satisfy $\epsilon_{\rm HO}$($q$)=$\epsilon_{\rm LU}$($q+1$). $\epsilon_{\rm LU}$($q+1$) is the lowest unoccupied Kohn-Sham eigenvalue in a positively charged defect system, in this case associated with the hole polaron  (Fig.~\ref{Fig.1 polaron diagram}b), and $\epsilon_{\rm HO}$($q$) is the highest occupied Kohn-Sham eigenvalue upon re-addition of an electron to the system with polaronic lattice distortion. We determine the value of $\alpha$ for the hole polaron by enforcing the condition $\epsilon_{\rm HO}$($q=0$)=$\epsilon_{\rm LU}$($q=1$). In various solids, enforcing this condition has shown to reproduce band gaps and localize polarons in agreement with state-of-the-art $GW$ computations \cite{miceli_nonempirical_2018,deak_choosing_2017,sadigh_variational_2015,chen_nonunique_2022,wing_role_2020}.

\begin{figure}[t]
 	\centering
 	\includegraphics[width=0.42\textwidth]{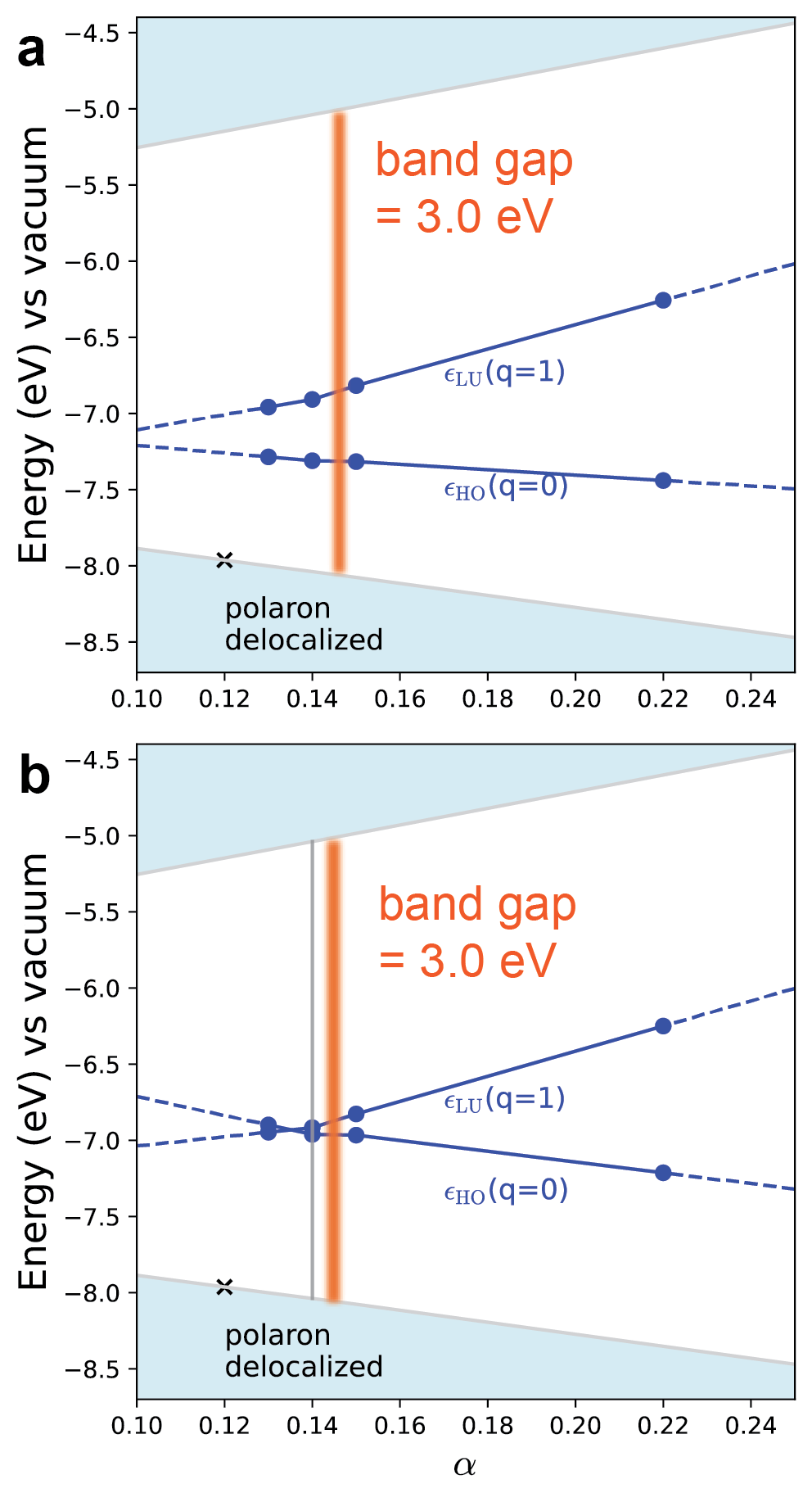}    
 	\caption{Single particle defect levels as a function of mixing parameter $\alpha$ for the hole polaron in rutile TiO$_{\rm 2}$ before and after finite size corrections. (a) shows the levels without finite size corrections, and (b) is after corrections are applied. The small polaron cannot be stablilized when alpha is below 12\% exchange. We find that with 14\% Fock exchange Koopmans' condition is satisfied. The corresponding pristine band gap is 3.0 eV. 
 	}
 	\label{Fig.2 KS}
\end{figure}

Figure~\ref{Fig.2 KS} shows how the band edges and the single particle level introduced by the polaron (when occupied or unoccupied) changes with $\alpha$. We first show the results without finite-size corrections in Fig. \ref{Fig.2 KS}a. As $\alpha$ increases the band gap opens up as expected and experimental band gap of 3.0 eV is reproduced at $\alpha=15\%$. The polaron single particle level also changes with $\alpha$ but they never cross. Satisfying the Koopmans' condition would require them to cross ($\epsilon_{\rm HO}$($q=0$)=$\epsilon_{\rm LU}$($q=1$)). It is notable that for the conditions satisfying the experimental band gap (3.0 eV), the Koopmans' condition is not satisfied for the polaron. Hole polarons cannot be stabilized below 12\% exchange (i.e., the localized solution cannot be found and the hole stays delocalized) and we have indicated this by dashed lines which are linear extrapolations. Using this extrapolation, it appears that the generalized Koopmans' condition should be satisfied at a low value of $\alpha$ where the polaron does not form (Fig. \ref{Fig.2 KS}a). This led previous authors to report that the hole polaron should be unstable \cite{elmaslmane_first-principles_2018}. 

The results in Figure \ref{Fig.2 KS}a however does not take into account finite size corrections. Under periodic boundary conditions, the long-range Coulomb interaction associated with a charged defect in a supercell causes spurious effects on defect formation energies and vertical transition energies. or vertical energies, it is necessary to take into account the finite-size corrections for both charged and neutral supercells. In particular, the neutral system can be subject to strong finite-size effect due to ionic polarization \cite{falletta_finite-size_2020,gake_finite-size_2020}. This can be corrected either by computing supercells of increasing size and extrapolating to the dilute limit, which is computationally intensive, or by applying a model correction scheme. We use the scheme proposed by Falletta, Wiktor, and Pasquarello \cite{falletta_finite-size_2020}, which is highly accurate in correcting vertical transitions (see SI). Figure \ref{Fig.2 KS}b plots the evolution of the polaron energy level including the finite size correction and show a significant difference, where a small fraction of exchange can stabilize the polaron and satisfy Koopmans' condition. Using the PBE0 functional and charge corrections, we find that the Koopmans' condition is satisfied at $\alpha$ = 0.14 for the hole polaron. In particular, the same $\alpha$ fulfilling the Koopmans' condition also reproduces the experimental band gap of the rutile TiO$_2$ (3.0 eV). Charge corrections provides definitely a more physically consistent picture and will be applied to all energies and eigenvalues reported from here.

\begin{figure}[ht]
 	\centering
 	\includegraphics[width=0.45\textwidth]{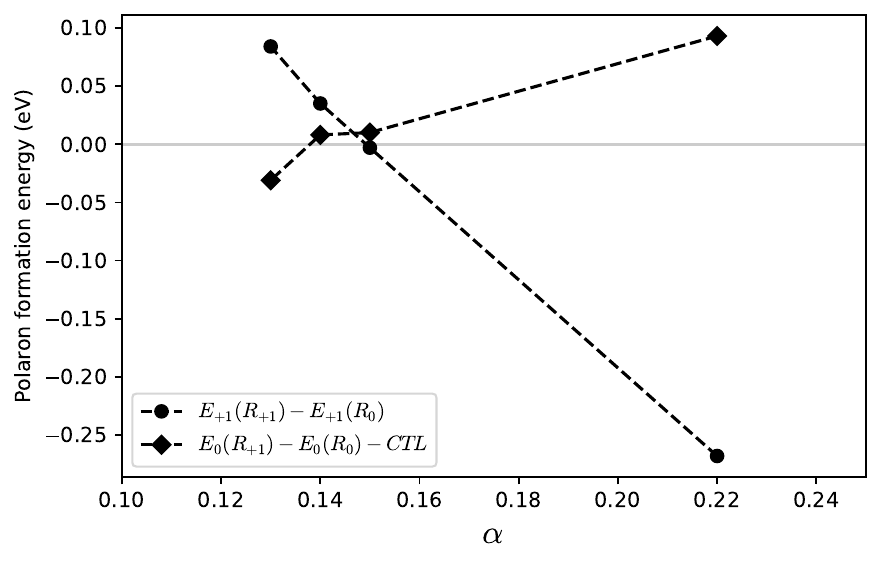}    
 	\caption{Polaron binding energy as a function of mixing parameter $\alpha$ computed two ways. Neither method of computing binding energy is constant with $\alpha$. At $\alpha$=0.14 eV, the hole polaron is thermodynamically unstable, with a binding energy of +0.01 eV. 
 	}
 	\label{Fig.3 binding energy}
\end{figure}

We now turn to computing the hole polaron binding energy. Hole polaron binding energy can be computed from positively charged computations as the total energy difference between the localized and delocalized hole (eq. \ref{charged_binding_energy}), or from neutral computations using the relationship between polaron binding energy, lattice energy, and ionization energy as depicted in Figure \ref{Fig.1 polaron diagram} (eq. \ref{neutral_binding_energy}). 
\begin{equation} \label{charged_binding_energy}
    E_{+1}(R_{+1}) - E_{+1}(R_0)  + E_\mathrm{corr}
\end{equation}
\begin{equation} \label{neutral_binding_energy}
    E_0(R_{+1}) - E_0(R_0) - CTL  + E_\mathrm{corr}
\end{equation}

where q=+1 is the charge of the polaron and CTL is the vertical charge transition level of the polaron (labeled as electronic energy in Fig \ref{Fig.1 polaron diagram}). Polaron binding energy is conventionally taken to be positive when the binding is favorable; however in this work we report favorable binding energies as negative to distinguish from unstable polaron binding energy.

The second equation has often been used in the literature to compute binding energy in an attempt to eliminate the need for charge corrections \cite{deak_polaronic_2011, kokott_first-principles_2018}; however, as discussed above, the vertical transition level has a large finite size effect and must be corrected. In addition, both of these methods are dependent on correctly choosing the fraction of Fock exchange, and are only coherent at Koopmans' compliant $\alpha$, as illustrated in Fig. \ref{Fig.3 binding energy}.

When using the Koopmans' compliant $\alpha$, we find that a localized hole polaron is higher in energy than the delocalized state by 10 meV. This indicates an unstable polaron and that the delocalized hole state is more stable than the small polaron in bulk rutile TiO$_2$.

\begin{table}[ht!]
    \caption{\label{Tab2: Literature} Selected theory publications on hole polaron in bulk rutile TiO$_{\rm 2}$.}
    \begin{ruledtabular}
    \begin{tabular}{c c c c c}
Publication & Functional & Polaron? & Binding energy & Ref \\ \hline 
 Deskins 2009   & GGA+U & Yes   & Not reported &
    \cite{deskins_intrinsic_2009}\\
 De\`ak 2011   & HSE06 & No    & Delocalized &
    \cite{deak_polaronic_2011}\\
 Cheng 2014   & HSE06 & Yes   & -0.57 eV &
    \cite{cheng_identifying_2014} \\
 Elmaslmane 2018   & PBE0($\alpha$)  & No    & Delocalized &
    \cite{elmaslmane_first-principles_2018}\\
 Kokott 2018   & HSE06 & No    & Unstable &
    \cite{kokott_first-principles_2018}\\
 Pada Sarker 2024   & GGA+U & Yes   & -0.56 eV &
    \cite{pada_sarker_prediction_2024}\\
 This work & PBE0($\alpha$)& No & Unstable&--
 \label{table_literature}
\end{tabular}

\end{ruledtabular}
\end{table}

Rutile TiO$_2$ hole polarons have been previously computed in the literature following several theoretical approaches. We present in Table \ref{Tab2: Literature} a selection of previous theoretical results on the rutile hole polaron. Our results show that the hole polaron is unstable, in agreement with \cite{deak_polaronic_2011,elmaslmane_first-principles_2018,kokott_first-principles_2018}. These studies all made some effort to use a Koopmans' compliant functional, but none applied corrections to the electronic levels. De\`ak et al. used the HSE06 functional and reported that the hole was delocalized \cite{deak_polaronic_2011}. However, we found that hole polarons are localized within HSE06. Elmaslmane et al. concluded that the hole polaron did not form, as the polaron did not satisfy Koopmans' condition at any $\alpha$, consistent with our computations without correction (see Fig. \ref{Fig.2 KS}a) \cite{elmaslmane_first-principles_2018}. Kokkott et al., using an approach that includes corrections for supercell size and self-interaction, briefly state that they do not find a stable hole polaron in rutile TiO$_2$, in agreement with our results \cite{kokott_first-principles_2018}.

Each study that reports a stable hole polaron overbinds the hole by using a very large $\alpha$ or Hubbard $U$ parameter. The most recent of these from Sarker et al.\ uses a Hubbard $U$ value of 10 eV applied to the oxygen 2$p$-orbital at the polaron site, a similar approach to the one used by Deskins \cite{pada_sarker_prediction_2024, deskins_intrinsic_2009}. This $U$-value was chosen to reproduce the polaron level from HSE06, which overbinds holes. We tested the use of GGA+$U$ to model the hole polaron in rutile TiO$_2$. DFT+$U$ is commonly used to study polarons as it is more computationally efficient than hybrid DFT. However, we could not reproduce similar binding energies and polaron configurations to Koopmans' compliant hybrid DFT using a Hubbard $U$ applied on oxygen. A comparison between DFT+$U$ and hybrid functionals is provided in Table S1.

We only report the bulk polaron here. Surface polarons can and do form on TiO$_2$ rutile. However, our work indicates that charge correction and Koopmans' compliant functionals could be important for the correct theoretical treatment of these surface polarons.

\begin{figure}[ht]
 	\centering
 	\includegraphics[width=0.4\textwidth]{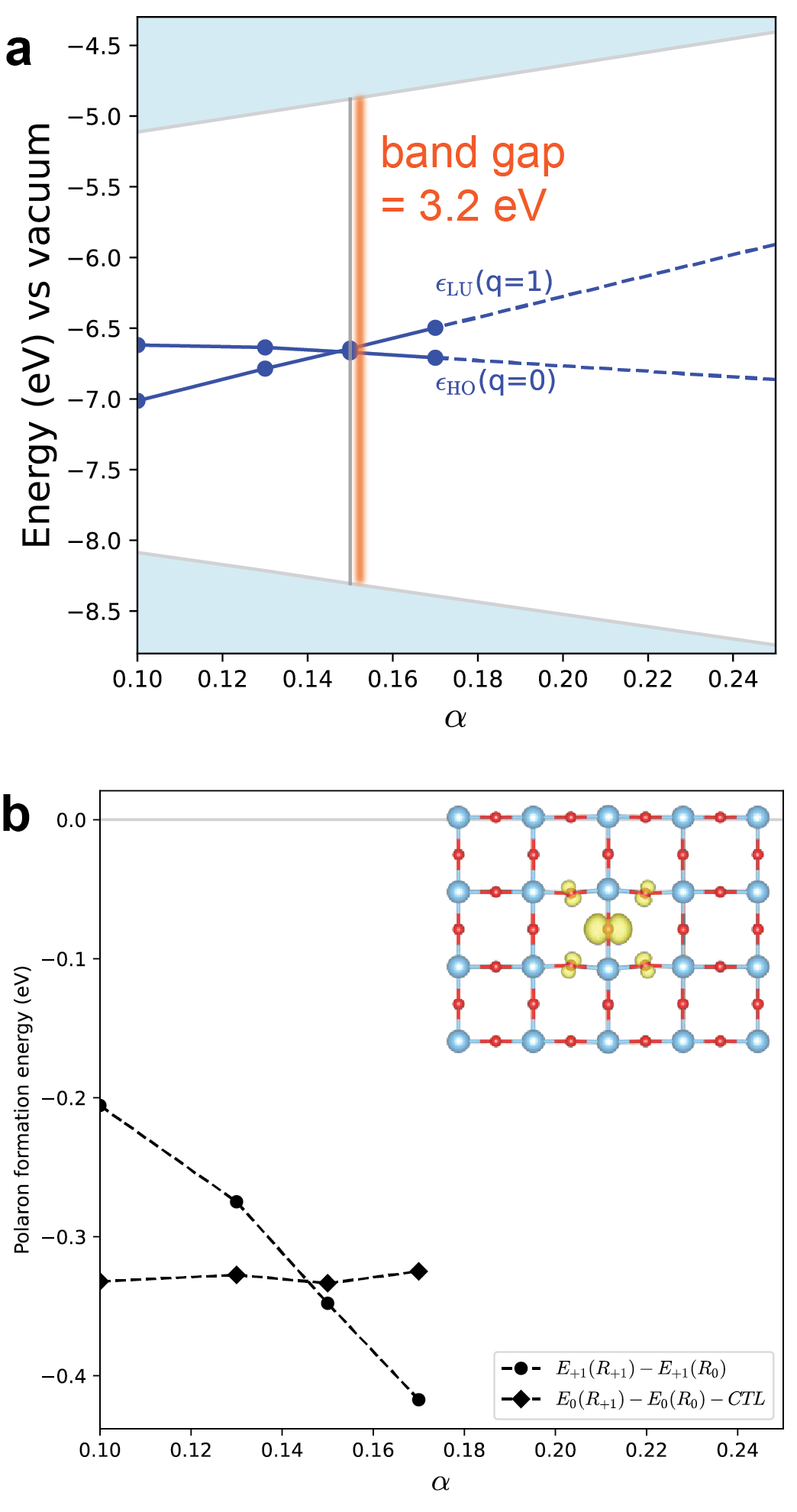}    
 	\caption{a) Single particle defect levels as a function of mixing parameter $\alpha$ for the hole polaron in anatase TiO$_{\rm 2}$ b) Computed polaron binding energy in anatase as a function of mixing parameter $\alpha$. Inset shows spin density contour of the hole polaron.}
 	\label{Fig.4 anatase}
\end{figure}\textbf{}

As a benchmark of our approach, we use the same procedure for the hole polaron in anatase TiO$_{\rm 2}$. Anatase is a known polymorph of TiO$_{\rm 2}$ where there is broad consensus on the existence of hole polarons, which have been reported both computationally and experimentally \cite{elmaslmane_first-principles_2018, deak_quantitative_2012, park_direct_2022}. We find that $\alpha$ of 0.15 leads to a Koopmans' compliant polaron level and a correct band gap (Figure \ref{Fig.4 anatase}a).  The charge transition level is 1.4 eV and a binding energy of $-0.32$ eV is computed (Figure \ref{Fig.4 anatase}.b).

A stable hole polaron agrees with theory and experiment. The binding energy of 320 meV and vertical transition level of 1.4 eV agrees with previous theoretical work \cite{deak_quantitative_2012, elmaslmane_first-principles_2018}. This level represents an upper bound on the experimental midgap state, since it does not include temperature or non-adiabatic effects - the level observed experimentally is closer to 0.7 eV above the valence band \cite{park_direct_2022}. 

From theory, the polaron is stable across a wide range of exchange fractions, so carefully checking Koopmans' condition is less important here; however it remains important for accurate estimations of polaron mid-gap level and hopping energy. The stability of the small hole polarons in anatase has thus been easier to determine than in rutile.  


In summary, the existence and stability of small hole polarons in bulk rutile TiO$_2$ has been contentious with contradicting theoretical and experimental evidences. Using a hybrid functional that is tuned to enforce Koopmans' condition and charge corrections, we have modeled the hole polaron in TiO$_2$. We conclude that the small hole polaron is unstable in bulk rutile and lies 10 meV above the delocalized hole. Applying the same metholodogy for anatase TiO$_2$ confirms on the contrary the existence of a stable bulk small hole polaron in anatase. Our work clarifies this controversial question and stresses the importance of Koopmans' compliance and charge correction in the modeling of small polarons.

\section{Computational Methods}

Hybrid DFT computations were performed in the CP2K package using the truncated PBE0($\alpha$) functional with a cutoff of 6 angstroms \cite{kuhne_cp2k_2020, vandevondele_quickstep_2005, vandevondele_efficient_2003, guidon_robust_2009}. Polarons were modeled using a 4x4x6 supercell for rutile and a 4x4x2 supercell for anatase. Only the $\Gamma$-point was sampled in reciprocal space. We used  Geodecker-Tuter-Hutter pseudopotentials \cite{krack_pseudopotentials_2005}. The primary basis set was triple-$\zeta$ or double-$\zeta$ quality with valence and polarization \cite{vandevondele_gaussian_2007}. Hybrid computations were accelerated with auxiliary density matrix method \cite{guidon_auxiliary_2010}. For the pristine configuration, atomic positions and volume were relaxed until all ionic forces were under 0.01 eV/angstrom for each $\alpha$. Polaron configurations were determined by distorting the lattice around an oxygen atom, removing one electron, and allowing the atoms to relax in fixed volume until the forces on all atoms was less than 0.01 eV/angstrom. 

The dielectric constants were calculated with the Perdew–Burke–Ernzerhof (PBE) functional in VASP using density functional perturbation theory (DFPT). Finite size corrections were applied following Falleta et. al \cite{falletta_finite-size_2020}.

\section{Acknowledgments}
\begin{acknowledgments}
This work was supported by the U.S. Department of Energy, Office of Science, Office of Basic Energy Sciences under Award No. DE-SC0023415 (the Center for Electrochemical Dynamics for Reactions on Surfaces, an Energy Frontier Research Center.) The computational work used resources of the National Energy Research Scientific Computing Center (NERSC), a Department of Energy Office of Science User Facility using NERSC award BES-ERCAP0028800.
\end{acknowledgments}

\bibliography{main}
\end{document}


\title{Supplementary information: Do Small Hole Polarons form in Bulk Rutile TiO$_2$?}

\author{Shay McBride} 
\affiliation{Thayer School of Engineering, Dartmouth College, Hanover, New Hampshire 03755, USA}
\author{Wei Chen}
\affiliation{Institute of Condensed Matter and Nanoscicence (IMCN),
Universit\'{e} catholique de Louvain,
Louvain-la-Neuve 1348, Belgium}
\author{Tanja \'Cuk}
\affiliation{University of Colorado, Boulder, Colorado 80309, USA}
\author{Geoffroy Hautier} 
\affiliation{Thayer School of Engineering, Dartmouth College, Hanover, New Hampshire 03755, USA}

\date{\today}

\maketitle
\clearpage

\section*{Supplementary note 1: Finite size corrections}

As mentioned in the text, the neutral polaron computation eigenvalues and energies show significant size effects. The correction scheme we use is effective at correcting this. We use computed dielectric constants $\varepsilon_\infty$ = 8.3 and $\varepsilon_0$ = 550 for rutile, and $\varepsilon_\infty$ = 6.8 and $\varepsilon_0$ = 43.5 for anatase.

\begin{figure*}[ht]
 	\centering
 	\includegraphics[width=0.6 \textwidth]{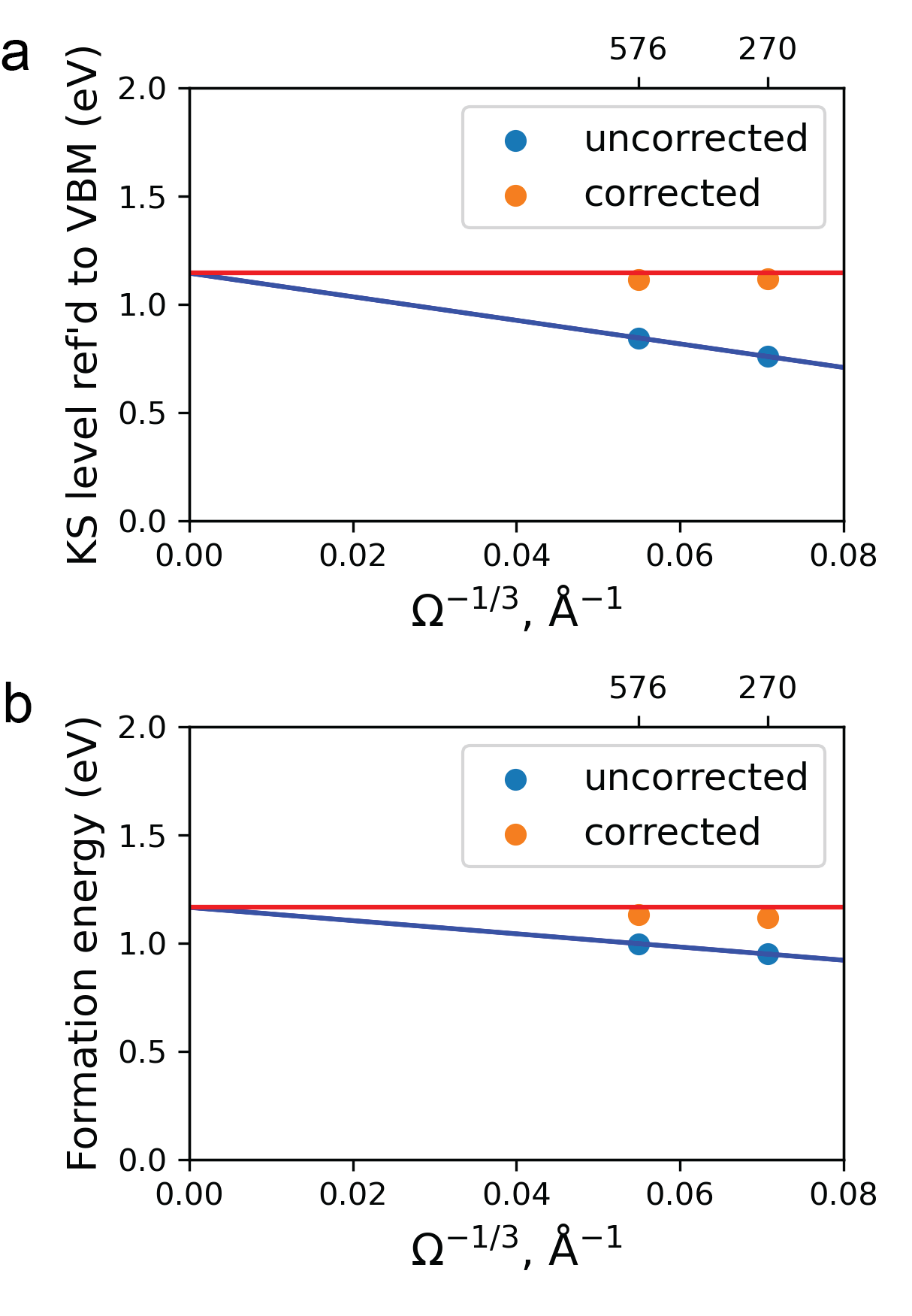}    
 	\caption{Scaling with inverse supercell size ($\Omega^{-1/3}$) for rutile TiO2 of (a) the KS level  and (b) the formation energy of the neutral defect in the hole polaron configuration. Corrected and uncorrected values are shown. The dilute limit is shown as a red line.}
 	\label{Fig.S2 SI phonon}
\end{figure*}
\clearpage

\section*{Supplementary note 2: HSE06 and PBE+U}

We report polaron formation energy and vertical transition levels using a Hubbard U value applied to oxygen 2p orbitals. We find that while using U=10 eV gives similar results to the HSE06 functional, both this U value and the HSE06 functional overbind the hole. We did not find a U-value that gave similar results to the PBE0($\alpha=0.14$) functional, as the hole delocalized below U=8 eV.

\begin{figure}[ht]
    \centering
    \begin{minipage}{0.8\textwidth}
        \centering
        \captionof{table}{Polaron binding energy (E$_{pol}$), computed as energy difference between hole polaron and delocalized hole, and KS levels of neutral and charged defect in hole polaron configuration, referenced to the pristine valence band maximum, for DFT+U and selected hybrid functionals.}

        \begin{ruledtabular}
        \begin{tabular}{cccc}
            U on O$_{2p}$ (eV) & E$_{pol}$ (eV) &  $\epsilon_{LU}$(q=+1) & $\epsilon_{HO}$(q=0) \\ \hline
            6 & delocalized & -- & --\\
            8 & -0.31 & 1.4 & 0.9 \\
            10 & -0.5 &2.0 & 1.2 \\
            PBE$_0(\alpha=0.14)$  &+0.04 & 1.1 & 1.1\\
            HSE06 (CP2K) &-0.2 & 1.7 & 1.4\\
            HSE06 (VASP) &-0.2 & 1.7 & 1.4
        \end{tabular}
        \end{ruledtabular}
    \end{minipage}\hfill
    \label{Table SI: HubbardU}
\end{figure}